\documentclass{article}
\usepackage{spconf,amsmath,graphicx}
\usepackage{epsfig}
\usepackage{amsmath}
\usepackage{amssymb}
\usepackage{color}
\usepackage{url}

\usepackage{caption}
\usepackage{subcaption}
\usepackage{float}

\def\0{{\mathbf 0}}
\def\1{{\mathbf 1}}

\def\x{{\mathbf x}}

\def\A{{\mathbf A}}

\def\I{{\mathbf I}}

\def\P{{\mathbf P}}

\def\X{{\mathbf X}}

\def\ie{{\textit{i.e.}}}

\def\cF{{\mathcal F}}
\def\cG{{\mathcal G}}

\def\bPhi{{\boldsymbol \Phi}}
\def\bPsi{{\boldsymbol \Psi}}

\title{Learned Nonlinear Predictor for Critically Sampled \\ 3D Point Cloud Attribute Compression}

\name{
    Tam Thuc Do$^{\ast}$, Philip A. Chou$^{\dag}$, Gene Cheung$^{\ast}$
}
\address{
\small\begin{minipage}{\linewidth}\begin{center}
\begin{tabular}{ccc}
$^{\ast}$York University & \hspace*{0.5in} & $^{\dag}${\tt packet.media} \\
Toronto, Canada && Seattle, USA 
\end{tabular}
\end{center}\end{minipage}
}

\begin{document}

%
\maketitle
\begin{abstract}
We study 3D point cloud attribute compression via a volumetric approach: assuming point cloud geometry is known at both encoder and decoder, parameters $\theta$ of a continuous attribute function $f: \mathbb{R}^3 \mapsto \mathbb{R}$ are quantized to $\hat{\theta}$ and encoded, so that discrete samples $f_{\hat{\theta}}(\x_i)$ can be recovered at known 3D points $\x_i \in \mathbb{R}^3$ at the decoder.
Specifically, we consider a nested sequences of function subspaces $\mathcal{F}^{(p)}_{l_0} \subseteq \cdots \subseteq \mathcal{F}^{(p)}_L$, where $\mathcal{F}_l^{(p)}$ is a family of functions spanned by B-spline basis functions of order $p$, $f_l^*$ is the projection of $f$ on $\mathcal{F}_l^{(p)}$ represented as low-pass coefficients $F_l^*$, and $g_l^*$ is the residual function in an orthogonal subspace $\mathcal{G}_l^{(p)}$ (where $\mathcal{G}_l^{(p)} \oplus \mathcal{F}_l^{(p)} = \mathcal{F}_{l+1}^{(p)}$) represented as high-pass coefficients $G_l^*$.
In this paper, to improve coding performance over \cite{do2023volumetric}, we study predicting $f_{l+1}^*$ at level $l+1$ given $f_l^*$ at level $l$ and encoding of $G_l^*$ for the $p=1$ case (RAHT($1$)).
For the prediction, we formalize RAHT(1) linear prediction in MPEG-PCC in a theoretical framework, and propose a new nonlinear predictor using a polynomial of bilateral filter.
We derive equations to efficiently compute the critically sampled high-pass coefficients $G_l^*$ amenable to encoding.
We optimize parameters in our resulting feed-forward network on a large training set of point clouds by minimizing a rate-distortion Lagrangian. 
Experimental results show that our improved framework outperforms the MPEG G-PCC predictor by $11\%$--$12\%$ in bit rate.
\end{abstract}

\vspace*{-0.5ex}
\section{Introduction}
\label{sec:intro}
We study the compression of \textit{point clouds}---a popular 3D imaging representation for a wide range of applications such as virtual reality and immmersive teleconferencing \cite{Zhang2013, dEonHMC:2017}.
In particular, assuming geometry (\ie, point cloud locations $\x_i \in \mathbb{R}^3$) is known both at encoder and decoder, we focus on point cloud attribute compression: quantize and encode parameters $\theta$ to $\hat{\theta}$ that specify a volumetric attribute function $f: \mathbb{R}^3 \mapsto \mathbb{R}$, so that discrete samples $f_{\hat{\theta}}(\x_i)$ can be recovered at the decoder. 
Following \cite{do2023volumetric}, we consider a nested sequence of function subspaces $\cF_{l_0}^{(p)} \subseteq \ldots \subseteq \cF_L^{(p)}$, where $\cF_l^{(p)}$ is a family of functions spanned by B-spline basis functions of order $p$.
We encode orthonormalized \textit{low-pass coefficients} $\bar F_l^*$ of the projection $f_l^*$ of $f$ on $\cF_l^{(p)}$ and orthonormalized \textit{high-pass coefficients} $\bar G_l^*$ of the projection $g_l^*$ of $f$ on the orthogonal complement subspace $\cG_l^{(p)}$, where $\cG_l^{(p)} \oplus \cF_l^{(p)} = \cF_{l+1}^{(p)}$.
See \cite{do2023volumetric} for details of the coding framework.

In this paper, to further improve coding performance over \cite{do2023volumetric}, we investigate predicting $f_{l+1}^*$ at level $l+1$ given $f_l^*$ at level $l$ and encoding of $G_l^*$ for the $p=1$ case.
Specifically, for the prediction we formalize \textit{Region Adaptive Hierarchical Transform} RAHT(1) linear prediction in MPEG-PCC \cite{QueirozC:16, ChouKK:20} in a theoretical framework, and propose a new nonlinear (pseudo-linear) predictor using a polynomial of a \textit{bilateral filter} \cite{tomasi98, antonio2013polybf}.
To efficiently encode high-pass coefficients $G_l^*$, we derive equations to compute critical samples for compact encoding, contrasting with non-critically sampled (redundant) high-pass coefficient $G_l^*$ computed in \cite{do2023volumetric}.

Data-driven image compression, leveraging the recent explosion in deep neural network development, has become prevalent \cite{BalleEtAl:20,mentzer2020high,hu2021learning}, with success extended to 3D point cloud geometry compression as well \cite{guarda2020deep,tang2020deep,Quach2020ImprovedDP}. 
However, on the task of point cloud attribute compression, the proposed learned neural methods \cite{ShengLLXLW:21,IsikCHJT:21,WangM:22} have yet to outperform MPEG G-PCC.  
The reason may be traced to the fundamental difficulty with point clouds that they are sparse, irregular, and changing.  
The usual \textit{dense} space-invariant convolutions used by neural networks for image compression as well as point cloud geometry compression become \textit{sparse} convolutions when the attributes are known at only specific positions.  
Sparse convolutions (convolutions that treat missing points as having attributes equal to zero) should be normalized, but until now it has not been clear how.
In our approach, we ``unroll" our orthonormalized B-spline-based coding framework into an interpretable feed-forward network that is amenable to data-driven end-to-end optimization of sparse parameters, similar to \textit{algorithm unrolling} in the literature \cite{li2019algorithm, monga2021algorithm}.
Experimental results show noticeable coding gain over \cite{do2023volumetric} and MPEG RAHT(1) with low complexity.

\vspace*{-0.5ex}
\section{Coding Framework}
\label{sec:framework}
Denote by $f:\mathbb{R}^3\mapsto\mathbb{R}$ a volumetric function defining an attribute field over $\mathbb{R}^3$.
Then, a point cloud can be seen as samples of $f$, with attributes $y_i=f(\mathbf{x}_i)$, where $\x_i \in \mathbb{R}^3$ is sample $i$'s 3D position. 
In \cite{GPCC:21, SchwarzEtAl:18}, first-order RAHT ($p=1$) is used as the core transform to code the attributes given the positions. 
Then, in \cite{ChouKK:20, do2023volumetric}, the transform is generalized to $p$-th-order RAHT($p$).  In RAHT($p$), the function $f$ is projected onto a nested sequence of \textit{function spaces} $\mathcal{F}_{l_0}^{(p)}\subseteq\cdots\subseteq\mathcal{F}_L^{(p)}$. 
Each function space $\mathcal{F}_l^{(p)}$ is a parametric family of functions spanned by a set of $N_l$ volumetric B-spline basis functions $\phi_{l,\mathbf{n}}^{(p)}(\mathbf{x})=\phi_{0,\mathbf{0}}^{(p)}(2^l\mathbf{x}-\mathbf{n})$ of order $p$ and scale $l$ for offsets $\mathbf{n}\in\mathcal{N}_l\subset\mathbb{Z}^3$.
The projections are the \textit{best approximations} $f_l^*$ of $f$ at various levels of detail $l$. 
The difference or {\em residual function} $g_l^* = f_{l+1}^* - f_l^*$ between consecutive approximations can be shown to lie in the function space $\mathcal{G}_l^{(p)}$ defined as the orthogonal complement of $\mathcal{F}_l^{(p)}$ in $\mathcal{F}_{l+1}^{(p)}$.
Then (dropping the superscript $p$ for brevity), one may write
\begin{eqnarray}
\cF_L & = &\cF_{l_0} \oplus \cG_{l_0+1} \oplus \cdots \oplus \cG_l \oplus \cdots \cG_{L-1}   \;\;\; \mbox{and}
\label{eq:sum_of_subspaces} \\
f_L^* & = & f_{l_0}^* + g_{l_0+1}^* + \cdots + g_l^* + \cdots + g_{L-1}^* .
\label{eqn:fg0_gL}
\end{eqnarray}
Thus, assuming $f_L^*(\mathbf{x}_i)=f(\mathbf{x}_i)$ for all $\mathbf{x}_i$, one may code $f$ by coding $f_{l_0}^*$, $g_{l_0+1}^*,\ldots,g_{L-1}^*$, which are each represented as coefficients (in an agreed upon basis) to be quantized and entropy-coded. 
In previous works, \cite{do2023volumetric} showed how to compute these coefficients efficiently in a feed-forward neural network. We describe this next.

\subsection{Low-pass Coefficients}
\label{sec:low_pass_coefs}

Let $\bPhi_l=[\phi_{l,\mathbf{n}}]$ be the row-vector of basis functions at level $l$, and $\bPhi_l^\top=[\langle\phi_{l,\mathbf{n}},\cdot\rangle]$ be the column-vector of functionals that are inner products with the basis functions at level $l$. 
Then, $\bPhi_l^\top f=[\langle\phi_{l,\mathbf{n}},f\rangle]$ is the column-vector of inner products of the basis functions with $f$, and $\bPhi_l^\top\bPhi_l$ is the matrix of inner products of the basis functions with themselves, namely the \textit{Gram matrix}.  
Here, the inner product is given by $\langle g,f\rangle=\sum_ig(\mathbf{x}_i)f(\mathbf{x}_i)$.
Under this inner product, we may restrict attention to the finite number $N_l$ of basis functions $\phi_{l,\mathbf{n}}$
for which there exists at least one point $\mathbf{x}_i$ where $\phi_{l,\mathbf{n}}(\mathbf{x}_i)\neq0$,
i.e., whose support contains at least one point in the point cloud.
 
Denote by $f_l^*$ the projection of the function $f$ onto the function subspace $\mathcal{F}_l$, and let $f_l^*$ be represented by the column vector of {\em low-pass coefficients} $F_l^*$ in the basis $\bPhi_l$.
The low pass coefficients can be calculated by solving the \textit{normal equation}:
\begin{equation} \label{eq:normal_equation_F}
F_{l}^* = (\bPhi_{l}^{\top} \bPhi_{l})^{-1} \bPhi_{l}^{\top} f .
\end{equation}
Assume that there is a sufficiently high resolution $L$ such that the $i$-th basis function $\phi_{L,\mathbf{n}_i}$ is 1 on $\mathbf{x}_i$ and 0 on $\mathbf{x}_j$ for $j\neq i$, namely $\phi_{L,\mathbf{n}_i}(\mathbf{x}_j)=\delta_{ij}$. 
Then, at the highest level of detail $L$, we have $\bPhi_L^\top\bPhi_L=\I$, $\bPhi_L^\top f=[y_i]$, so that the vector of low-pass coefficients is given by 
\begin{equation}
    F_L^*=(\bPhi_L^\top\bPhi_L)^{-1}\bPhi_L^\top f=[y_i].
    \label{eq:low_pass_L}
\end{equation}
Then, denote by $\A_{l}=[a_{ij}^l]$ the $N_{l} \times N_{l + 1}$ matrix whose $i$th row expresses the $i$-th basis function $\phi_{l,\mathbf{n}_i}$ of $\mathcal{F}_l$ in terms of the basis functions $\phi_{l+1,\mathbf{m}_j}$ of $\mathcal{F}_{l+1}$, namely
\begin{equation}
\bPhi_l = \bPhi_{l+1} \A_l^\top .
\label{eqn:Phi_ell_in_terms_of_Phi_ell1}
\end{equation}
Using (\ref{eqn:Phi_ell_in_terms_of_Phi_ell1}), starting with $\bPhi_L^\top\bPhi_L=\I$, for each $l<L$,
the Gram matrix at level $l$ can be computed from the Gram matrix at level $l+1$ as
\begin{eqnarray}
\bPhi_l^\top \bPhi_l & = & \A_l\bPhi_{l+1}^\top \bPhi_{l+1} \A_l^\top .
\label{eqn:PhiTPhi}
\end{eqnarray}
Further, starting with $F_L^*=[y_i]$ in \eqref{eq:low_pass_L}, for each $l<L$, the low-pass coefficients can be expressed
\begin{align}
F_l^* & = (\bPhi_l^\top \bPhi_l)^{-1} \bPhi_l^\top f \label{eqn:f_star}\\
& = (\bPhi_l^\top \bPhi_l)^{-1} \A_l \A_{l+1}\cdots \A_{L-1}\bPhi_L^\top f .
\label{eqn:F_ell_star_from_F_ell1_star}
\end{align}
This can be computed easily by defining $\tilde{F}_l^* = (\bPhi_{l}^\top \bPhi_{l}) F_{l}^*$, which we call the vector of {\em un-normalized} low-pass coefficients.  Then, starting from $\tilde{F}_L^* = (\bPhi_{L}^\top \bPhi_{L}) F_{L}^*=F_L^*$, for each $l<L$,
$\tilde{F}_l^*$
can be calculated using \eqref{eqn:analysis_lowpass}, and $F_l^*$ can be calculated using \eqref{eqn:normalization}:
\begin{align}
    \tilde{F}_l^* &= \A_l \tilde{F}_{l+1}^* \label{eqn:analysis_lowpass}\\
    F_l^* &= (\bPhi_{l}^\top \bPhi_{l})^{-1} \tilde{F}_l^* \label{eqn:normalization}.
\end{align}

Defining $\mathbf{d}_{ij} \triangleq \mathbf{m}_j-2\mathbf{n}_i$ with $\mathbf{m}_j \in \mathcal{N}_{l+1}$ and $\mathbf{n}_j \in \mathcal{N}_{l}$, we parameterize $\A_{l}$ by a vector of size $2\times 2 \times 2 = 8$ as
\begin{equation}
a_{ij}^l = \left\{\begin{array}{ll}
w_{\mathbf{d}_{ij}} & \mbox{if}~~ \mathbf{d}_{ij} \in \{0, 1\}^3 \\ 
0 & \mbox{otherwise}
\end{array}\right . .
\label{eq:RAHTp}
\end{equation}
RAHT($1$) is a special case of this parameterization, in which $w_{\mathbf{d}_{ij}}=1$. 
It can be seen that $\mathbf{A}_l$ is a space-invariant but {\em sparse} and {\em disjoint} convolutional matrix. 
With this form of $\A_l$, \eqref{eqn:analysis_lowpass} is an ordinary sparse convolution, and \eqref{eqn:normalization} is a trivial normalization, since the matrix $\bPhi_l^\top \bPhi_l$ is diagonal.

\subsection{High-pass Coefficients}
\label{sec:High-pass-coefficients}

As introduced in \cite{do2023volumetric}, it is possible to represent the residual function $g_l^*$ in a basis $\bPsi_l$ for $\mathcal{G}_l$.
Denote by $G_l^*$ a column-vector of {\em high-pass} coefficients corresponding to the row-vector of basis functions $\bPsi_l$, both of length $N_{l+1}-N_l$, such that $g_l^*=\bPsi_l G_l^*$.  The coefficients $G_l^*$ can be calculated via the normal equation
\begin{equation} 
G_{l}^* = (\bPsi_{l}^{\top} \bPsi_{l})^{-1} \bPsi_{l}^{\top} g_l^* .
\label{eq:normal_equation_G}
\end{equation}
By definition, $\mathcal{G}_l$ is orthogonal to $\mathcal{F}_l$ and $\mathcal{G}_l\subset\mathcal{F}_{l+1}$. 
Hence, $\bPsi_l$ can be written as a linear combination of $\bPhi_{l+1}$, and its inner product with $\bPhi_l$ is zero, namely 
\begin{align}
    \bPsi_{l} = \bPhi_{l+1} \mathbf{Z}_l^\top \;\mbox{and}\; \bPhi_{l}^\top \bPsi_{l} = \mathbf{0}. \label{eqn:psi_l_condition}
\end{align}
There are many choices for $\mathbf{Z}_l$; we will derive one as follows.
Define a \textit{selection matrix} $\mathbf{I}_l^{b}$ of size $({N}_{l+1} - {N}_l) \times {N}_{l+1}$ that selects a subset $b$ of basis vectors from level $l+1$,
\begin{equation}
    \mathbf{\Phi}_{l+1}^b = \mathbf{\Phi}_{l+1} \mathbf{I}_l^{b\top}.
\end{equation}
Then, define $\mathbf{Z}_l^\top$ as 
\begin{eqnarray}
    \mathbf{Z}_l^\top &=& \left [ \mathbf{I} - \mathbf{A}_l^\top (\mathbf{\Phi}_{l}^\top\mathbf{\Phi}_{l})^{-1}\mathbf{A}_l
    (\mathbf{\Phi}_{l+1}^\top\mathbf{\Phi}_{l+1}) \right ] \mathbf{I}_l^{b\top} \\
    &=& \mathbf{I}_l^{b\top} - \mathbf{A}_l^\top (\mathbf{\Phi}_{l}^\top\mathbf{\Phi}_{l})^{-1}(\mathbf{\Phi}_{l}^\top\mathbf{\Phi}_{l+1}^b)
    \label{eq:Z_operation}
\end{eqnarray}
so that \eqref{eqn:psi_l_condition} are satisfied:
\begin{align}
    \mathbf{\Phi}_{l}^\top&\mathbf{\Psi}_{l} = \mathbf{\Phi}_{l}^\top\mathbf{\Phi}_{l+1} \mathbf{Z}_l^\top \\
    &= (\mathbf{\Phi}_{l}^\top\mathbf{\Phi}_{l+1}^b) - (\mathbf{\Phi}_{l}^\top\mathbf{\Phi}_{l})(\mathbf{\Phi}_{l}^\top\mathbf{\Phi}_{l})^{-1}(\mathbf{\Phi}_{l}^\top\mathbf{\Phi}_{l+1}^b) \\
    &= 0 .
\end{align}
Because of the form of $\mathbf{A}_l$, the choice of $\mathbf{I}_l^{b}$ is also trivial: 
simply drop one point for each disjoint block $\{0, 1\}^3$, and select leftover points.
This guarantees {\em critical sampling}, i.e., the number coefficients is exactly equal to the number of input points.

With the basis $\bPsi_{l} = \bPhi_{l+1} \mathbf{Z}_l^\top$ for $\mathcal{G }_l^*$ constructed, we first define $\delta F_{l+1}^*$ as {\em residual coefficients} that represent the residual function $g_l^*=f_{l+1}^*-f_l^*$ in the basis of $\bPhi_{l+1}$ for $\mathcal{F}_{l+1}$:
\begin{equation}
    \delta F_{l+1}^* = (\bPhi_{l+1}^\top\bPhi_{l+1})^{-1}\bPhi_{l+1}^\top g_l^* = F_{l+1}^*-\A_l^\top F_{l}^* .
    \label{eqn:delta_F}
\end{equation}
Then, any function expressed as $\delta f_{l+1} = \bPhi_{l+1} \delta F_{l+1}$, for example $g^*_l = \bPhi_{l+1} \delta F_{l+1}^*$, can be projected onto $\mathcal{G }_l$ and represented by coefficients $G_l$ in the basis $\bPsi_{l}$ as follows:
\begin{align}
    G_l &= (\bPsi_{l}^{\top} \bPsi_{l})^{-1} \bPsi_{l}^{\top} \delta f_{l+1} \\
    &= (\mathbf{Z}_l \bPhi_{l+1}^{\top} \bPhi_{l+1} \mathbf{Z}_l^\top )^{-1} \mathbf{Z}_l (\bPhi_{l+1}^{\top}\bPhi_{l+1}) \delta F_{l+1} ,
    \label{eq:orthor_proj_G_L}
\end{align}
where $(\bPhi_{l+1}^{\top}\bPhi_{l+1})$, $\mathbf{Z}_l^\top$, and $\mathbf{Z}_l = (\mathbf{Z}_l^\top)^\top$ are defined in \eqref{eqn:PhiTPhi} and \eqref{eq:Z_operation}.
Note that $\bPsi_l^\top\bPhi_l=\mathbf{0}$ implies $\mathbf{Z}_l (\bPhi_{l+1}^{\top}\bPhi_{l+1}) \mathbf{A}_l^\top=\mathbf{0}$, and hence from \eqref{eqn:delta_F} and \eqref{eq:orthor_proj_G_L} we also have
\begin{equation}
    G_l^* = (\mathbf{Z}_l \bPhi_{l+1}^{\top} \bPhi_{l+1} \mathbf{Z}_l^\top )^{-1} \mathbf{Z}_l (\bPhi_{l+1}^{\top}\bPhi_{l+1}) F_{l+1}^* .
    \label{eq:orthor_proj_G_L_from_F_lp1}
\end{equation}

As done in \cite{do2023volumetric}, instead of quantizing the coefficients $F^*_{l_0}$ or $G^*_l$ directly, we quantize the \textit{orthonormalized} coefficients calculated as
\begin{align}
    \bar{F}_{l_0}^* &
    = (\bPhi_{l_0}^{\top} \bPhi_{l_0})^{\frac{1}{2}} F_{l_0}^* = (\bPhi_{l_0}^{\top} \bPhi_{l_0})^{-\frac{1}{2}} \tilde{F}_{l_0}^*\label{eqn:orthor_normal_equation_F} \;\;\;\mbox{and} \\
    \bar{G}^*_{l}  &
    = (\bPsi_{l}^{\top} \bPsi_{l})^{\frac{1}{2}} G^*_l \\
    &=  (\bPsi_{l}^{\top} \bPsi_{l})^{-\frac{1}{2}}  \mathbf{Z}_l (\bPhi_{l+1}^{\top}\bPhi_{l+1}) \delta F_{l+1}^* .
\label{eq:orthor_normal_equation_G}
\end{align}
To compute \eqref{eqn:orthor_normal_equation_F} and \eqref{eq:orthor_normal_equation_G}, we implement the operator $\X^{-\frac{1}{2}}$ as a $P$-layer feed-forward network with coefficients initially derived from the $P$-th order Taylor expansion of $x^{-\frac{1}{2}}$. Later on, these Taylor coefficients become trainable parameters in the training process. Thus, the orthonormalized coefficients will correspond to the orthonormal basis functions $\bar{\bPhi}_{l_0} = \bPhi_{l_0} (\bPhi_{l_0}^\top \bPhi_{l_0})^{-\frac{1}{2}}$ and $\bar{\bPsi}_l = \bPsi_l (\bPsi_l^\top \bPsi_l)^{-\frac{1}{2}}$, which are then quantized for entropy coding.

Noting that $\bPhi_{l+1}F_{l+1}^* = f_{l+1}^* = f_l^* + g_l^* = \bPhi_lF_l^* + \bPsi_lG_l^*$, we have $\bPhi_{l+1}^\top\bPhi_{l+1}F_{l+1}^* = \bPhi_{l+1}^\top\bPhi_lF_l^* + \bPhi_{l+1}^\top\bPsi_lG_l^* = \bPhi_{l+1}^\top\bPhi_{l+1}\mathbf{A}_l^\top F_l^* + \bPhi_{l+1}^\top\bPhi_{l+1}\mathbf{Z}_l^\top G_l^*$, or
\begin{equation}
    F^*_{l+1} = \mathbf{A}_l^\top F_l^* + \mathbf{Z}_l^\top G_l^* .
\end{equation}
Thus, once the quantized orthonormalized coefficients $\hat{\bar{F}}_{l_0}^*$ and $\hat{\bar{G}}_l^*$ for $l=l_0,\ldots,L-1$ are recovered, and converted to quantized coefficients $\hat{F}_{l_0}^*=(\bPhi_{l}^{\top} \bPhi_{l})^{-\frac{1}{2}}\hat{\bar{F}}_{l_0}^*$ and $\hat{G}_l^*=(\bPsi_{l}^{\top} \bPsi_{l})^{-\frac{1}{2}}\hat{\bar{G}}_l^*$ for $l=l_0,\ldots,L-1$, we recover for $l=l_0,\ldots,L-1$
\begin{equation}
\hat F_{l+1}^* = \mathbf{A}_l^\top \hat F_l^* + \mathbf{Z}_l^\top \hat G_l^* .
\label{eq:allpass_reconstruction}
\end{equation}
Finally $\hat f_L^*=\bPhi_L\hat F_L^*$ is the reconstruction of $f$.

\vspace*{-1.5ex}
\section{Constrained Prediction \& Residual Coding}
\label{sec:implement}


    

\begin{figure*}[t]
    \noindent\makebox[\textwidth]{
    \begin{subfigure}{0.35\textwidth}
        \centering
        \includegraphics[width=\textwidth]{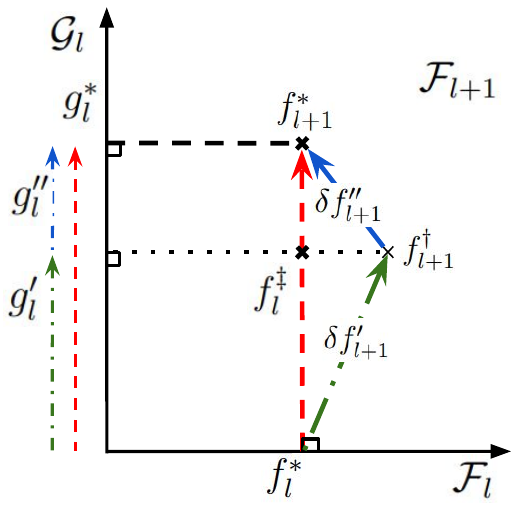}
        \caption{}
        \label{fig:adjusted_up_sampling}
    \end{subfigure} 
    \hspace{-0.2cm}
    \begin{subfigure}{0.7\textwidth}
        \centering
        \includegraphics[width=\textwidth, trim=2.5cm 0.2in 3.7in 2.7in,clip]{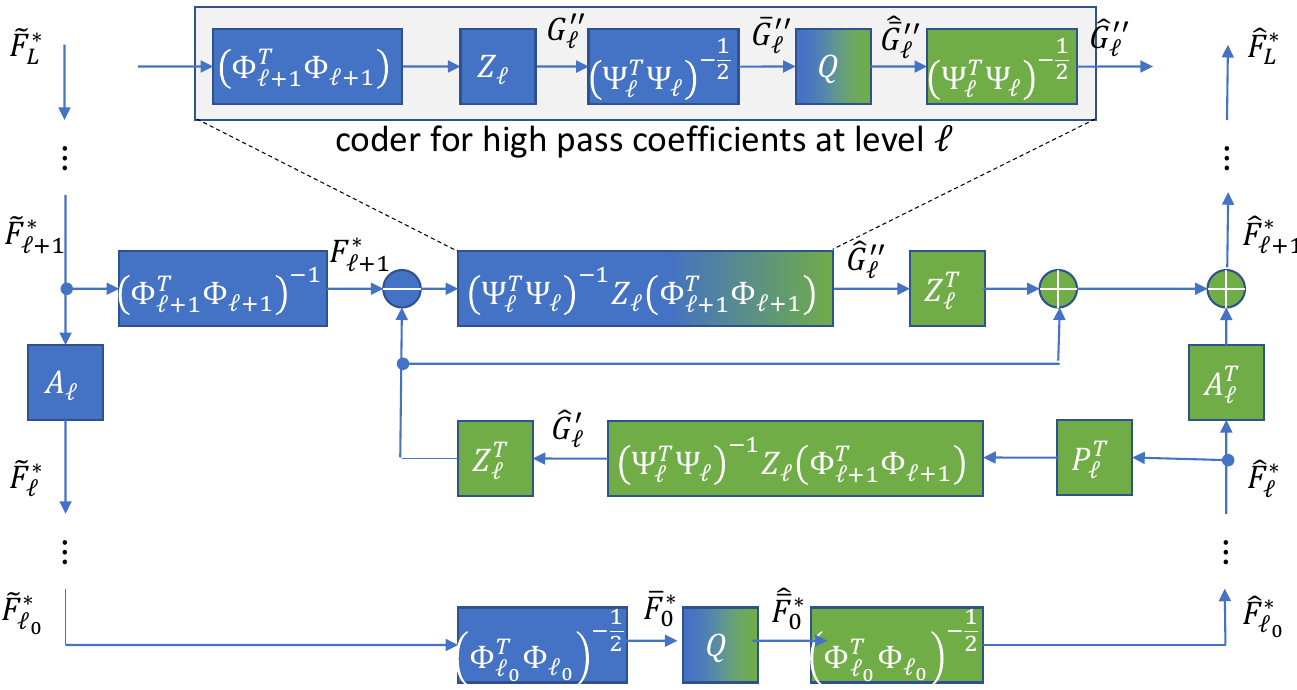}
        \caption{}
        \label{fig:feedforward_network_block_diagram}
    \end{subfigure}}
    \caption{(a) Visualization of constrained Up-sampling Prediction, (b) Multilevel feedforward network implementing point cloud attribute encoder (blue and green) and decoder (green).}
    \label{fig:code_gain_performance}
\end{figure*}

\subsection{Constrained Up-sampling Prediction}

In this work, we reduce the magnitude of the residual coefficients by generalizing the 
synthesis operation  $\A_l^\top$ to a more general $2\times$ upsampling prediction  $\P_l^\top$. In particular, we define the \textit{upsampling predictor}, or simply the {\em predictor}, at level $l$ as
\begin{equation}
     f_{l+1}^{\dagger} = \text{predictor}( f_{l}^*) = \bPhi_{l+1} \P_l^\top  F_l^* .
\end{equation}
Then, we define $\delta f'_{l+1} \triangleq f_{l+1}^{\dagger} - f_{l}^*$ and $\delta f''_{l+1} \triangleq  f_{l+1}^* - f_{l+1}^{\dagger}$, and let $g'_{l}$ and $g''_{l}$ be the projections of $\delta f'_{l+1}$ and $\delta f''_{l+1}$ onto $\mathcal{G}_l$, respectively. 
We can now write
\begin{equation}
    f_{l+1}^* =  f_{l}^* + g^ *_{l} = f_{l}^* + g'_{l} + g''_{l} = f_{l}^\ddagger + g''_{l} ,
\end{equation}
where $f_{l}^\ddagger = f_{l}^* + g'_{l}$ is the \textit{constrained prediction}, that is, the prediction constrained to lie in the coset $f_l^*+\mathcal{G}_l$.
The residual coefficients that express $\delta f'_{l+1}$ and $\delta f''_{l+1}$ in the basis $\bPhi_{l+1}$ can be written as
\begin{align}
    \delta F'_{l+1}
    & = (\bPhi_{l+1}^\top\bPhi_{l+1})^{-1}\bPhi_{l+1}^\top (f_{l+1}^{\dagger} - f_{l}^*) \\
    & = \P_l^\top F_l^* - \mathbf{A}_l^\top F_{l}^* \\ 
    \delta F_{l+1}''
    & = (\bPhi_{l+1}^\top\bPhi_{l+1})^{-1}\bPhi_{l+1}^\top
     (f_{l+1}^* - f_{l+1}^{\dagger}) \\
    & = F_{l+1}^* - \P_l^\top F_l^* .
\end{align}
Then, coefficients that represent $g'_{l}$ and $g''_{l}$ are calculated by applying the projection operation (using eq.~\eqref{eq:orthor_proj_G_L}) introduced in Section\;\ref{sec:High-pass-coefficients} as follows: 
\begin{align}
    G_l' &= (\bPsi_{l}^{\top} \bPsi_{l})^{-1} \mathbf{Z}_l (\bPhi_{l+1}^{\top}\bPhi_{l+1})  \delta F'_{l+1} \\
    &=
    (\bPsi_{l}^{\top} \bPsi_{l})^{-1} \mathbf{Z}_l (\bPhi_{l+1}^{\top}\bPhi_{l+1}) \P_l^\top F_l^* ,
    \label{eq:G_l_one}
    \\
    G''_l &= (\bPsi_{l}^{\top} \bPsi_{l})^{-1} \mathbf{Z}_l (\bPhi_{l+1}^{\top}\bPhi_{l+1}) \delta F_{l+1}'' \\
    &=
    G_l^* - G_l' .
    \label{eq:G_l_two}
\end{align}
Thus, $G_l'$ predicts the high-pass coefficients $G_l^*$, and $G_l''$ is the residual of this prediction, which is orthonormalized, quantized, and encoded.
Analogous to \eqref{eq:allpass_reconstruction}
we recover $F_{l+1}^*$ for $l=l_0,\ldots,L-1$ by applying \eqref{eq:G_l_one} to $\hat F_l^*$ to get $\hat G_l'$ 
\begin{eqnarray}
\hat F_{l+1}^*
& = &
\mathbf{A}_l^\top \hat F_l^* + \mathbf{Z}_l^\top \hat G_l^* \\
& = & \mathbf{A}_l^\top \hat F_l^* + \mathbf{Z}_l^\top (\hat G_l' + \hat G_l'') .
\end{eqnarray}

The intuition behind this is illustrated in Fig.\;\ref{fig:adjusted_up_sampling}.
The constrained prediction $f_{l+1}^\ddagger$ is better than the unconstrained prediction $f_{l+1}^\dagger$ because the prediction residual is smaller: $||g_l''||\leq||\delta f_{l+1}''||$.
Fig.\;\ref{fig:feedforward_network_block_diagram} shows the implementation as a feed-forward network. 
Next, we discuss the choice of $\P_l^\top$ such that this can lead to coding gain.

\subsection{Locally Linear Prediction} \label{sec:locally_linear_enhancement}

Inspired by MPEG G-PCC \cite{SchwarzEtAl:18, Wang2023predGPCC, GPCC:21}, we define  $\P_l^\top$---the \textit{up-sampling prediction} operation---as a locally linear operator. 
In vector form, we express it as two separate operations:
\begin{equation}
    \P_l^\top F_{l}^* = \mathbf{D}^{-1} \mathbf{W} \mathbf{A}_l^\top F_{l}^*
    \label{eq:matrix_LLWA}
\end{equation}
where $\mathbf{A}_l^\top$ is the up-sampling operation, followed by weighted convex combination of coefficients in the 27-neighborhood of node-$i$, $\mathcal{M}(i) = \{j | \mathbf{k}_{ij}=(\mathbf{n}_i - \mathbf{n}_j) \in \{-1, 0, 1\}^3 \}$, $\mathbf{W} = [ w_{\mathbf{k}_{ij}}]_{N\times N}$ is the matrix of weights, and $\mathbf{D}$ is the diagonal degree matrix with diagonal elements $\mathbf{D}_{ii} = \sum_j w_{\mathbf{k}_{ij}} $. 
Hence, the weights are space-invariant, so we parametrize $\mathbf{W}$ by 27 parameters, which are trained later.

\subsection{Polynomial Bilateral Filter Prediction}

In addition, we propose another prediction scheme by replacing locally linear $ \mathbf{D}^{-1} \mathbf{W}$ with a polynomial of \textit{bilateral filter} (PBF) (introduced in \cite{antonio2013polybf}). 
Specifically, we calculate the weights $w_{\mathbf{k}_{ij}}$ in a signal-dependent manner:
\begin{equation}
w_{\mathbf{k}_{ij}} = \exp \left( -\frac{\| \mathbf{n}_i-\mathbf{n}_j \|^2}{2\sigma_x^2} \right) \, \exp \left( -\frac{\| \mathbf{x}[i]-\mathbf{x}[j] \|^2}{2\sigma_y^2} \right) ,
\end{equation}
where $\sigma_x$ and $\sigma_y$ are the filter parameters, which are trainable.
In matrix form,
\begin{equation}
    \text{BF}(\mathbf{x}) = \mathbf{D}_\mathbf{x}^{-1} \mathbf{W}_\mathbf{x} \mathbf{x} .
    \label{eq:matrix_BF}
\end{equation}
This dependence makes the BF both space-varying and nonlinear in $\mathbf{x}$, though because of its form, it is often termed {\em pseudo-linear}. To obtain the PBF, first we define $\mathbf{L}_\mathbf{x} \triangleq \mathbf{D}_\mathbf{x} - \mathbf{W}_\mathbf{x}$ as the combinatorial Laplacian matrix for the 27-neighborhood graph. We then interpret the bilateral filter \eqref{eq:matrix_BF} as a graph filter,
\begin{equation}
    \text{BF}(\mathbf{x}) = (\mathbf{I} - \boldsymbol{\mathcal{L}}_r) \mathbf{x} ,
\end{equation}
where $\boldsymbol{\mathcal{L}}_r = \mathbf{D}_\mathbf{x}^{-1} \mathbf{L}_\mathbf{x}$ is the random walk Laplacian matrix. The bilateral filter can then be generalized into a graph filter with spectral response function belonging to the class of polynomial functions with maximum degree $K$.
\begin{align}
    \text{PBF}(\mathbf{x}) = \textit{h}(\boldsymbol{\mathcal{L}}_r) \mathbf{x} &= r_0 \prod_{k=0}^{K} (\mathbf{I} - r_k \boldsymbol{\mathcal{L}}_r) \mathbf{x}.
    \label{eq:polynomials_BF}
\end{align}
The polynomials coefficients $\{ r_k\}_{k=0}^K$ are trained from data. 
As proved in \cite[Thm.~5.1]{antonio2013polybf}, operation $\textit{h}(\boldsymbol{\mathcal{L}}_r)$ can be implemented as a cascade of bilateral filters:
\begin{equation}
    (\mathbf{I} - r_k \boldsymbol{\mathcal{L}}_r) = (1-r_k)\mathbf{I}+r_k \mathbf{D}_\mathbf{x}^{-1} \mathbf{W}_\mathbf{x} .
\end{equation}
Thus, we have the following prediction:
\begin{align}
    \P_l^\top F_l^* = \text{PBF}(\A_l^\top F_l^* ) = h(\mathcal{L}_r)\A_l^\top F_l^* .
    \label{eq:GBF_pred_module}
\end{align}

\vspace*{-0.5ex}
\section{Experiments}
\label{sec:results}

\begin{figure*}[t]
    \noindent\makebox[\textwidth]{
    \begin{subfigure}{0.40\textwidth}
        \centering
        \includegraphics[width=\textwidth,trim=0.05in 0.0in 0.05in 0.0in,clip]{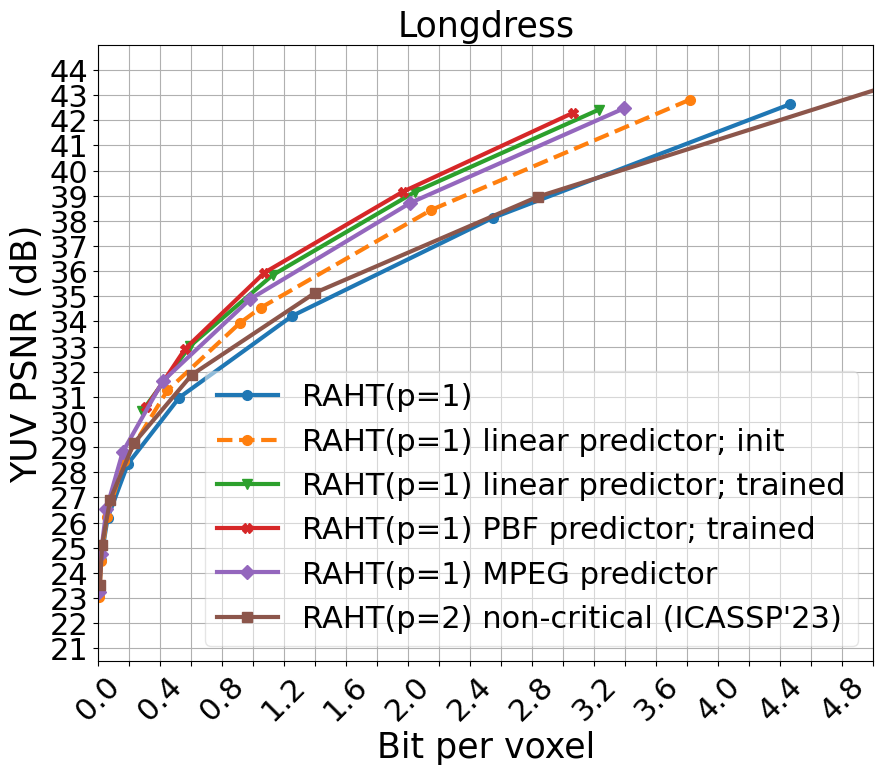}
        \caption{}
    \end{subfigure}
    \begin{subfigure}{0.41\textwidth}
        \centering
        \includegraphics[width=\textwidth,trim=0.05in 0.0in 0.05in 0.0in,clip]{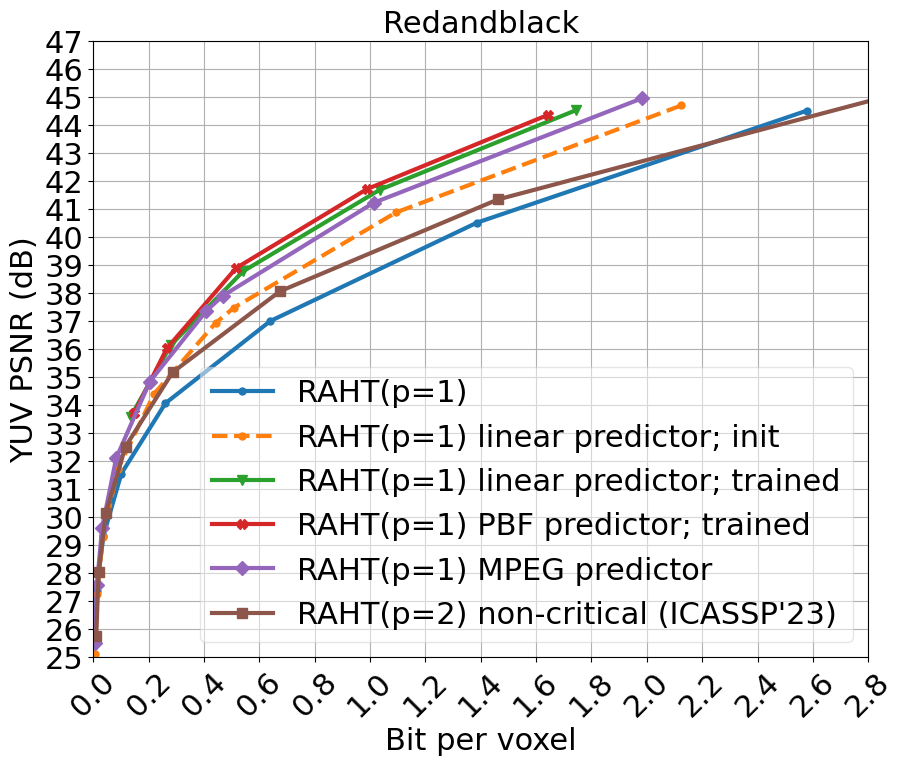}
        \caption{}
    \end{subfigure}}\\
    \noindent\makebox[\textwidth]{
    \begin{subfigure}{0.41\textwidth}
        \centering
        \includegraphics[width=\textwidth,trim=0.05in 0.0in 0.05in 0.0in,clip]{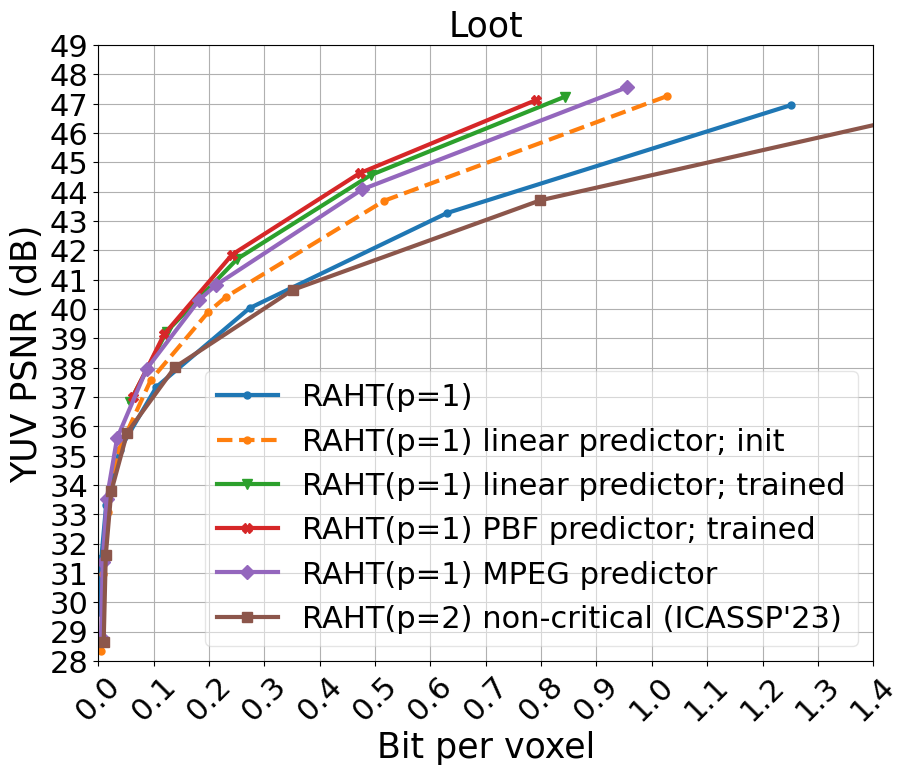}
        \caption{}
    \end{subfigure}
    \begin{subfigure}{0.40\textwidth}
        \centering
        \includegraphics[width=\textwidth,trim=0.05in 0.0in 0.05in 0.0in,clip]{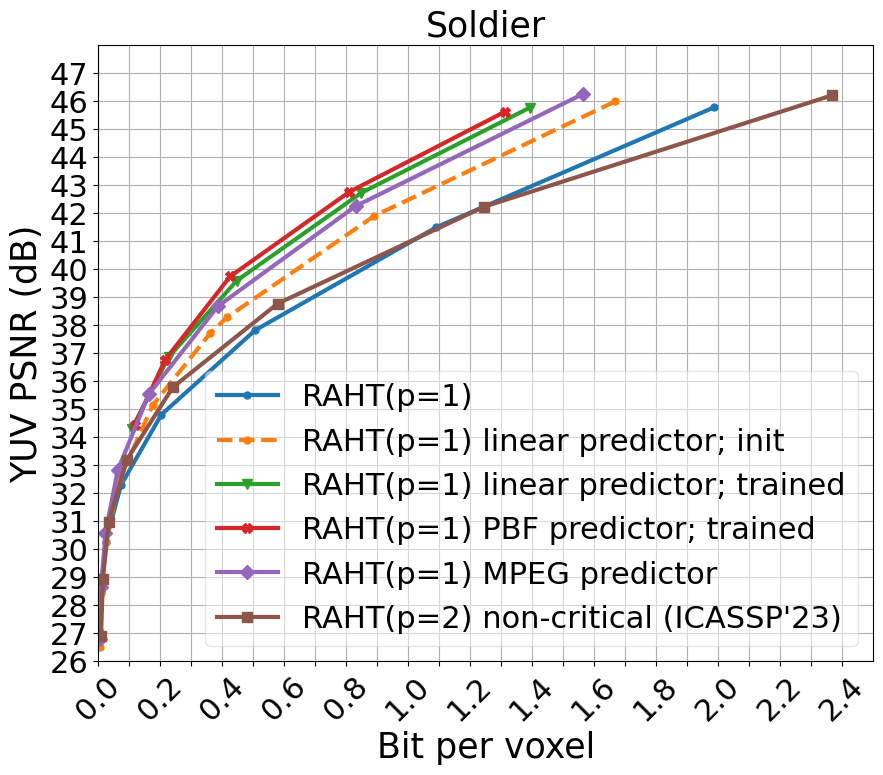}
        \caption{}
    \end{subfigure}}
    \caption{Rate-Distortion  curves: (a) \textit{Longdress}, (b) \textit{Redandblack}, (c) \textit{Loot}, (d) \textit{Soldier}}
    \label{fig:code_gain_performance}
\end{figure*}

\subsection{Training}

Our training dataset comprises voxelized point clouds computed from 3D models (meshes) publicly available at sketchfab\footnote{https://sketchfab.com/blogs/community/sketchfab-launches-public-domain-dedication-for-3d-cultural-heritage/}.
Our evaluation dataset comprises the MPEG point clouds \textit{Longdress}, \textit{Redandblack}, \textit{Loot}, and \textit{Soldier} \cite{dEonHMC:2017}.
All point clouds have 10-bit resolution, so their octrees have depth of $L=10$ and $N_L \sim  1e^6$.

For training, denote by $\Theta$ the set of trainable parameters, which for each level $l$ includes two kernels of size $2\times 2\times 2$ to parameterize $\mathbf{A}_l$ and $\mathbf{A}_l^\top$, two sets of $P+1$ polynomial coefficients to parameterize the sub-networks $(\bPsi_l^\top\bPsi_l)^{-1/2}$ for both encoder and decoder, 
and, depending on the choice of predictor, a $3\times 3\times 3$ kernel $\mathbf{W}$ for the linear predictor, or $\sigma_x$, $\sigma_y$, and $(K+1)$ polynomials coefficients $\{ r_k\}_{k=0}^K$ for the PBF predictor, as well as the parameters of the quantizer and entropy coder.

Our feed-forward network model minimizes the Lagrangian $J(\Theta) = D(\Theta) + \lambda R(\Theta)$, where $D(\Theta)$ is the mean of reconstruction square errors, $D(\Theta) = \frac{1}{N_L} \| F^*_L - \hat{F}^*_L \|_2^2$, and the Lagrange multiplier $\lambda$ is chosen to meet the bit rate constraint $R(\Theta) \leq R_0$. During training, we used the ``straight-through'' quantizer proxy $Q(\mathbf{x})=\mathbf{x}$ \cite{BalleCMSJAHT:21}, and the arithmetic coding proxy with coefficients $\{Y_l\}_{l=l_0}^{L}$: 
\begin{equation}
    R(\{Y_l\}_{l=l_0}^{L}, \Theta) = -\frac{1}{N_L} \sum_{l=l_0}^L \text{log}_2 p(Y_l; m_l, b_l, \Delta),
\end{equation}
where $\{Y_l\}_{l=l_0}^{L}$ are all the coefficients at each level and 
\begin{align}
    &p(Y_l; m_l, b_l, \Delta) \\
    &= \textit{CDF}_{m_l, b_l}\left (Y_l + \frac{\Delta}{2} \right) - \textit{CDF}_{m_l, b_l}\left (Y_l - \frac{\Delta}{2} \right).
    \vspace{-2ex}
\end{align}
with $\textit{CDF}_{m_l, b_l}$ was modeled by the Laplace distribution at each level $l$ with scalar location $m_l$, scalar diversity $b_l$, and a uniform quantization step size $\Delta$ as trainable parameters.

We implemented our model in Python using Jax. In all experiments, we picked $l_0=4$, $P=50$, $K=20$ and trained $\sim 900$ model parameters as described.
From $l_0=4$ to $L=10$ we only have 6 levels, so we can train on 6-bit ($64\times64\times64$) crops centered on random points per batch, instead using whole 10-bit resolution point clouds.
We used the Adam optimizer with learning rate of 0.0001 in all configurations.

For evaluation, we used the adaptive Run-Length Golomb-Rice (RLGR) entropy coder \cite{Malvar:2006} to calculate a realizable bit rate. 

\subsection{Baselines}
As a baseline, we estimated the performance of the transform and predictor in the MPEG geometry-based point cloud codec (G-PCC) \cite{SchwarzEtAl:18, Wang2023predGPCC, GPCC:21} by using RAHT($p=1$), which is the core transform of MPEG G-PCC, coupled with the MPEG G-PCC predictor. 
Each coefficient $F^*_{l+1}[\mathbf{m_j}]$ was linearly predicted from a neighborhood of coefficients, $F^*_l[\mathbf{n}_i]$, $\mathbf{n}_i\in\mathcal{M}(\lceil\mathbf{m}_j/2\rceil) \subset \mathcal{N}_l$ as
\begin{eqnarray}
     \P_l^\top F_{l}^*=   \mathbf{D}^{-1} \mathbf{W} F_{l}^{*},
\end{eqnarray}
where $\mathbf{W}$ was a $N_{l+1} \times N_{l} $ matrix with entries  $w_{\mathbf{m}_j, \mathbf{n}_i}=1/d(2\mathbf{n}_i+\mathbf{1},\mathbf{m}_j+\mathbf{1}/2)$ when $(\mathbf{m}_j - 2\mathbf{n}_i) \in \{ 0, 1 \}^2 + \{ -1, 0, 1 \}^3$ and zero otherwise, and $d$ was Euclidean distance.

\subsection{Results}

By placing the prediction component of G-PCC in a theoretical framework, we explored its potential using variety of prediction schemes (especially a nonlinear predictor under a training process). 
Hence, in this work, we limited the base transform to be RAHT($p=1$), so that we could take advantage of the simplicity of having a diagonal matrix $\bPhi_l^\top\bPhi_l$ and also critical sampling because the selection matrix $\mathbf{I}_l^{bT}$ is trivially determined.

For the prediction component, we test two different predictors introduced in Sec.~\ref{sec:implement}---\textit{Local linear} and \textit{Polynomial Bilateral Filter}---coupled with a training process to further improve prediction performance. 
In Fig.\;\ref{fig:code_gain_performance} we plot YUV PSNR as a function of bits per occupied voxel, where the coefficients of each color component (Y, U, V) are separately coded using RLGR.

As expected, with the freedom to adjust the predictor weights at different scales (or levels), the training process learns from data resulting in better prediction at each level individually. 
This is shown in Fig.\;\ref{fig:code_gain_performance}: starting from a reasonable initialization of the linear predictor (the dashed curves) the training process significantly improves the coding performance of the linear predictor (green curves) and surpasses the baseline MPEG G-PCC predictor by $5$ to $6\%$ in bit rate reduction. 

Next, we improve the prediction by incorporating signal dependency and non-linearity into the predictor (solid red curves). 
This results in an additional $6$ to $7\%$ bit rate reduction compared to the trained linear predictor (solid green curves), leading to an overall $11$ to $12\%$ bit rate improvement over the MPEG G-PCC prediction transform. 

Lastly, we compare our current results with non-critical RAHT($p=2$), a higher order B-spline transform with a non-critical residual coding method \cite{do2023volumetric}. 
We significantly outperform it due to the critical sampling property of the base transform RAHT($p=1$) coupled with a training process and non-linear prediction. 
However, we stress the possibility of further coding gain by using a higher order B-spline transform, albeit with higher complexity Gram matrix $\bPhi_l^\top\bPhi_l$ and non-trivially determined selection matrix $\mathbf{I}_l^{bT}$. We leave these aspects for future investigation.

\vspace*{-0.05in}
\section{Conclusion}
\label{sec:conclude}

To improve coding performance over a previously proposed 3D point cloud attribute compression framework \cite{do2023volumetric} based on a nested sequence of B-spline basis functions, we proposed two innovations: i) nonlinear prediction of low-pass coefficients at a higher level via a polynomial of a bilateral filter, and ii) efficient coding of critically sampled high-pass coefficients. 
We unrolled our framework into neural layers of a feed-forward network, whose parameters were trained using data by minimizing a rate-distortion Lagrangian.
Experiments show that our predictors outperformed the MPEG G-PCC predictor by $11$ to $12\%$ reduction in bit rate.

\bibliographystyle{IEEEbib}
\bibliography{refs}

\end{document}